\documentclass[twocolumn,twocolappendix]{aastex7}%

\usepackage{threeparttable}
\usepackage{color}
\usepackage{amsmath}
\usepackage{ulem}
\usepackage{hyperref}

\newcommand{\msun}{M_{\odot}}

\definecolor{ForestGreen}{RGB}{34,139,34}
\def\tw#1 {{\textcolor{ForestGreen}{#1}}\ }
\def\blue#1 {{\textcolor{blue}{#1}}\ }
\def\red#1 {{\textcolor{red}{#1}}\ }
\begin{document} 
\title{The Bigfoot: A footprint of a Coma cluster progenitor at $z = 3.98$}

\author{Hanwen Sun}
\affiliation{School of Astronomy and Space Science, Nanjing University, Nanjing 210093, China}
\affiliation{Key Laboratory of Modern Astronomy and Astrophysics, Nanjing University, Ministry of Education, Nanjing 210093, China}
\email{hanwensun@smail.nju.edu.cn}

\author[0000-0002-2504-2421,sname=Wang,gname=Tao]{Tao Wang}
\affiliation{School of Astronomy and Space Science, Nanjing University, Nanjing 210093, China}
\affiliation{Key Laboratory of Modern Astronomy and Astrophysics, Nanjing University, Ministry of Education, Nanjing 210093, China}
\email[show]{taowang@nju.edu.cn}

\author{Emanuele Daddi}
\affiliation{AIM, CEA, CNRS, Université Paris-Saclay, Université Paris
Diderot, Sorbonne Paris Cité, F-91191 Gif-sur-Yvette, France}
\email{emanuele.daddi@cea.fr}

\author{Qiaoyang Hao}
\affiliation{School of Astronomy and Space Science, Nanjing University, Nanjing 210093, China}
\affiliation{Key Laboratory of Modern Astronomy and Astrophysics, Nanjing University, Ministry of Education, Nanjing 210093, China}
\email{qiaoyanghao@smail.nju.edu.cn}

\author{Ke Xu}
\affiliation{School of Astronomy and Space Science, Nanjing University, Nanjing 210093, China}
\affiliation{Key Laboratory of Modern Astronomy and Astrophysics, Nanjing University, Ministry of Education, Nanjing 210093, China}
\email{kexu@smail.nju.edu.cn}

\author{David Elbaz}
\affiliation{AIM, CEA, CNRS, Université Paris-Saclay, Université Paris
Diderot, Sorbonne Paris Cité, F-91191 Gif-sur-Yvette, France}
\email{delbaz@cea.fr}

\author{Luwenjia Zhou}
\affiliation{School of Astronomy and Space Science, Nanjing University, Nanjing 210093, China}
\affiliation{Key Laboratory of Modern Astronomy and Astrophysics, Nanjing University, Ministry of Education, Nanjing 210093, China}
\email{wenjia@nju.edu.cn}

\author{Houjun Mo}
\affiliation{Department of Astronomy, University of Massachusetts, Amherst, MA 01003-9305, USA}
\email{hjmo@umass.edu}

\author{Huiyuan Wang}
\affiliation{School of Astronomy and Space Science, University of Science and Technology of China, Hefei, Anhui 230026, China}
\affiliation{Key Laboratory for Research in Galaxies and Cosmology, Department of Astronomy, University of Science and Technology of China, Hefei, Anhui 230026, China}
\email{whywang@ustc.edu.cn}

\author{Longyue Chen}
\affiliation{School of Astronomy and Space Science, Nanjing University, Nanjing 210093, China}
\affiliation{Key Laboratory of Modern Astronomy and Astrophysics, Nanjing University, Ministry of Education, Nanjing 210093, China}
\email{652023260002@smail.nju.edu.cn}

\author{Yangyao Chen}
\affiliation{School of Astronomy and Space Science, University of Science and Technology of China, Hefei, Anhui 230026, China}
\affiliation{Key Laboratory for Research in Galaxies and Cosmology, Department of Astronomy, University of Science and Technology of China, Hefei, Anhui 230026, China}
\email{yangyaochen.astro@foxmail.com}

\author{Shuowen Jin}
\affiliation{Cosmic Dawn Center (DAWN), Denmark}
\affiliation{DTU-Space, Technical University of Denmark, Elektrovej 327, DK2800 Kgs. Lyngby, Denmark}
\email{shuowen.jin@gmail.com}

\author{Yipeng Lyu}
\affiliation{AIM, CEA, CNRS, Université Paris-Saclay, Université Paris
Diderot, Sorbonne Paris Cité, F-91191 Gif-sur-Yvette, France}
\email{yipenglyu1998@gmail.com}

\author{Nikolaj Sillassen}
\affiliation{Cosmic Dawn Center (DAWN), Denmark}
\affiliation{DTU-Space, Technical University of Denmark, Elektrovej 327, DK2800 Kgs. Lyngby, Denmark}
\email{nbsi@space.dtu.dk}

\author{Kai Wang}
\affiliation{Institute for Computational Cosmology, Department of Physics, Durham University, South Road, Durham, DH1 3LE, UK}
\affiliation{Centre for Extragalactic Astronomy, Department of Physics, Durham University, South Road, Durham DH1 3LE, UK}
\email{wkcosmology@gmail.com}

\author{Tiancheng Yang}
\affiliation{School of Astronomy and Space Science, Nanjing University, Nanjing 210093, China}
\affiliation{Key Laboratory of Modern Astronomy and Astrophysics, Nanjing University, Ministry of Education, Nanjing 210093, China}
\email{652023260012@smail.nju.edu.cn}

   
\begin{abstract}
Protoclusters, galaxy clusters' high redshift progenitors, hold the keys to understanding the formation and evolution of clusters and their member galaxies. However, their cosmological distances and spatial extensions (tens of Mpc) have inhibited complete mapping of their structure and constituent galaxies, which is key to robustly linking protoclusters to their descendants. Here we report the discovery of the Bigfoot, a tridimensional structure at $z = 3.98$ including 11 subgroups traced by 55 (700) spectroscopic (photometric) redshifts with JWST, extending over $15\times 37$  $\times 49{\rm{cMpc^3}}$ in the PRIMER-UDS field. 
Bigfoot's large-scale and mass function of member galaxies closely match constrained simulations' predictions for the progenitors of today's most massive clusters  (${M_0} > 10^{15} \msun$). All subgroups with ${M_{\rm{h}}} > {10^{12.5}}{M_{_ \odot }}$ exhibit enhanced fractions of massive galaxies  ($>{10^{10.0} \msun}$) compared to lower-mass halos and field, demonstrating the accelerated formation of massive galaxies in massive halos. 
The presence of this massive protocluster with a large central halo (${10^{13.0} \msun}$) in a JWST deep field bears important cosmological implication that favors high ${\sigma _8}$ of PLANCK cosmology over low-redshift probes.
\end{abstract}


\section{introduction}
\label{sec:intro}

Massive galaxy clusters with virial masses $\log ({M_0}/{M_ \odot }) > 15$ are the largest gravitationally-bound systems in the local universe, whose central regions are dominated by the massive quiescent early-type galaxies \citep{Dressler1980}. How these massive systems and the member galaxies inside of them formed and evolved to their present state is still under debate. As suggested by both archaeology studies \citep{Thomas2010} and numerical simulations \citep[e.g.][]{Chiang2013,Henden2020,WangK2025}, most of the massive cluster galaxies are formed at $z>2$, making the progenitors of galaxy clusters (protoclusters) in the early universe important for understanding the formation scenario of galaxy clusters. 
In the last decade, many protoclusters at $z>2$ have been discovered using several methods, including searching for overdensities of galaxy number densities using the wide-field catalogs \citep[e.g.][]{Toshikawa2012, Toshikawa2018, Higuchi2019, Sillassen2022, Toshikawa2025}, detections of X-ray emissions from hot gas \citep[e.g.][]{Gobat2011, WangT2016}, overdensity of line emitters \citep[e.g.][]{Cai2017, Rubet2025}, overdensities of dusty star-forming galaxies \citep[e.g.][]{Miller2018, Oteo2018, Zhou2023, Sillassen2024, Foo2025}, and the groups around ultra-massive galaxies \citep[e.g.][]{McConachie2022, Hedge2025, Jespersen2025}.
However, due to the limited sensitivity and wavelength coverage of HST and ground-based telescopes, most of these protoclusters could only be studied with biased tracers like Lyman break galaxies (LBGs) and dusty star-forming galaxies (DSFGs), and only a small number of galaxy members could be confirmed with spectroscopic data. Without highly complete membership determinations, the global properties (e.g., total stellar masses) and future evolution path of these protoclusters remain elusive.

Recently, the James Webb Space Telescope (JWST) provides chances to discover \citep{Morishita2023, Helton2024, Jin2024, Jespersen2025, Li2025} and analyze \citep{PereaMartinez2024, SunF2024, Solimano2025, Umehata2025}  protoclusters at high redshifts with complete selection of the member galaxies down to the low stellar masses and with reliable estimation of their physical properties. However, due to the limited field of view of JWST, most of the JWST observations targeting protoclusters can only cover their core regions \citep{Arribas2024, Lamperti2024,Shimakawa2024}. In contrast, simulations suggest that the progenitors of massive cluster with $\log ({M_0}/{M_ \odot }) > 15$ can be highly extended, large-scale ($\sim$ 10000 cMpc$^3$) overdensities at cosmic dawn \citep{Chiang2013,WangHY2014,WangK2024}. In this case, only protoclusters located within JWST widest fields, such as J1001 at $z=2.51$ in COSMOS \citep{WangT2016,Sun2024} and the Cosmic Vine at $z=3.44$ in CEERS \citep{Jin2024}, could be studied with complete large-scale sampling to analyze their global properties and assess whether they will become local massive galaxy clusters. 

In this work, we report the discovery of protocluster PCL0217 (``the Bigfoot'') at $z = 3.98$. Located in the center of the PRIMER-UDS field with large JWST coverage from the PRIMER \citep{Dunlop2021} and the BEACON surveys \citep{Morishita2025}, the Bigfoot provides a rare opportunity to study the large-scale structure defined by a massive protocluster at cosmic dawn with mass-complete membership selection. These large-scale structures can be compared with the protoclusters predicted by constrained simulation \citep{WangHY2014} to study the future growth of the Bigfoot. 

This paper is organized as follows: Section~\ref{Sec:data} shows the data and the methods for sample selection; Section~\ref{sec:results} presents the results, including the stellar mass functions and density profiles of the Bigfoot; Section~\ref{sec:discussion} discusses the Bigfoot's implications on cosmology and the formation scenario of the first massive clusters. Throughout our work, we assume the cosmological model with ${H_0}{\rm{ = }}70~{\rm{km}} \cdot {{\rm{s}}^{ - 1}} \cdot {\rm{Mp}}{{\rm{c}}^{ - 1}}$, ${\Omega _{\rm{m}}} = 0.3$ and ${\Omega _\Lambda } = 0.7$. An initial mass function given by \citet{Kroupa2001} is used to estimate the stellar masses.

\section{Data and Sample construction}\label{Sec:data}

\subsection{Multi-wavelength catalog}
Our analyses for the Bigfoot are based on our UV-to-MIR multiwavelength deblending catalog for PRIMER-COSMOS and PRIMER-UDS, which are observed by the Public Release IMaging for Extragalactic Research (PRIMER, \citep{Dunlop2021}) survey with JWST/NIRCam and MIRI. Brief descriptions and validations of this catalog are available from \citet{WangT2025}, and more details will be shown in a forthcoming paper (H.Sun et al., in prep). In short, using the JWST images combining all available surveys in PRIMER-UDS and COSMOS, which are reduced by our custom-made pipeline based on the JWST Calibration Pipeline v1.13.4 \citep{Bushouse2024}, we perform source extraction on the stacked image of all JWST/NIRCam longwave (LW) bands. For each JWST-detected galaxy, we perform aperture photometry using APHOT \citep{Merlin2019} for the high-resolution data from JWST/NIRCam and HST, while the fluxes from low-resolution equipment, including JWST/MIRI, Spitzer, and ground-based telescopes, are measured by the template-fitting deblending photometry with TPHOT v2.0 \citep{Merlin2015,Merlin2016}. Lastly, we derive photometric redshifts using the EAZY software \citep{Brammer2008} for all JWST-detected sources in our photometric catalog. In this work, we use the data in PRIMER-UDS to study the Bigfoot, while the data in PRIMER-COSMOS can be used to investigate the field galaxies at the same redshift.

\subsection{Structure identification with spectroscopic data} 

\begin{figure*}[htbp]
\centering
\includegraphics[width=0.68\textwidth]{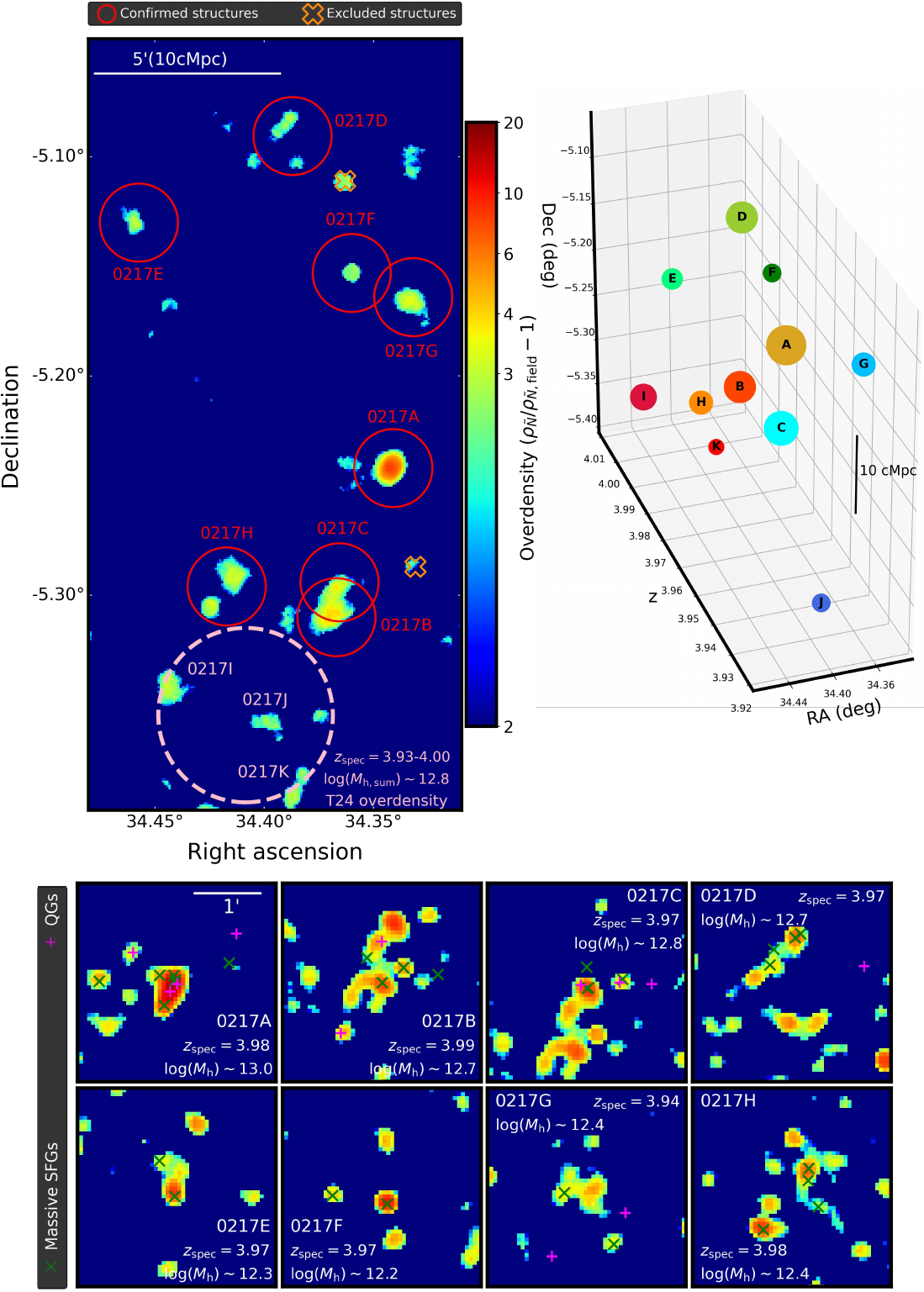}
\caption{\label{Fig:density}\footnotesize \textbf{The weighted number overdensity map of PCL0217--``the Bigfoot''.} The upper left panel shows the distribution of the weighted number overdensity of galaxies with $3.73 < {z_{{\rm{phot}}}} < 4.23$ in PRIMER-UDS. The red open circles show the position of the 8 confirmed subgroups observed by the JWST/PRIMER survey \citep{Dunlop2021}, the pink circle shows a previously reported overdensity of quiescent galaxies \citep{Tanaka2024} covered by the JWST/BEACON survey \citep{Morishita2025}. The upper right panel shows the 3-dimensional distribution of the 11 subgroups. An animated version of this figure is available at \href{https://box.nju.edu.cn/f/d6e29ce33a704cd29439/?dl=1}{https://box.nju.edu.cn/f/d6e29ce33a704cd29439/?dl=1}. The lower figures show the enlarged overdensity map of the eight confirmed subgroups covered by the PRIMER survey. The purple and red crosses show the location of massive (${M_{\rm{*}}} > {10^{9.5}}{M_{_ \odot }}$) SFGs and quiescent galaxies (QGs). The overdensities shown in the upper and lower panels are measured in circular areas with radius 1' and 15", respectively.}
\end{figure*}

\begin{figure*}[!htbp]
\centering
\includegraphics[width=0.9\textwidth]{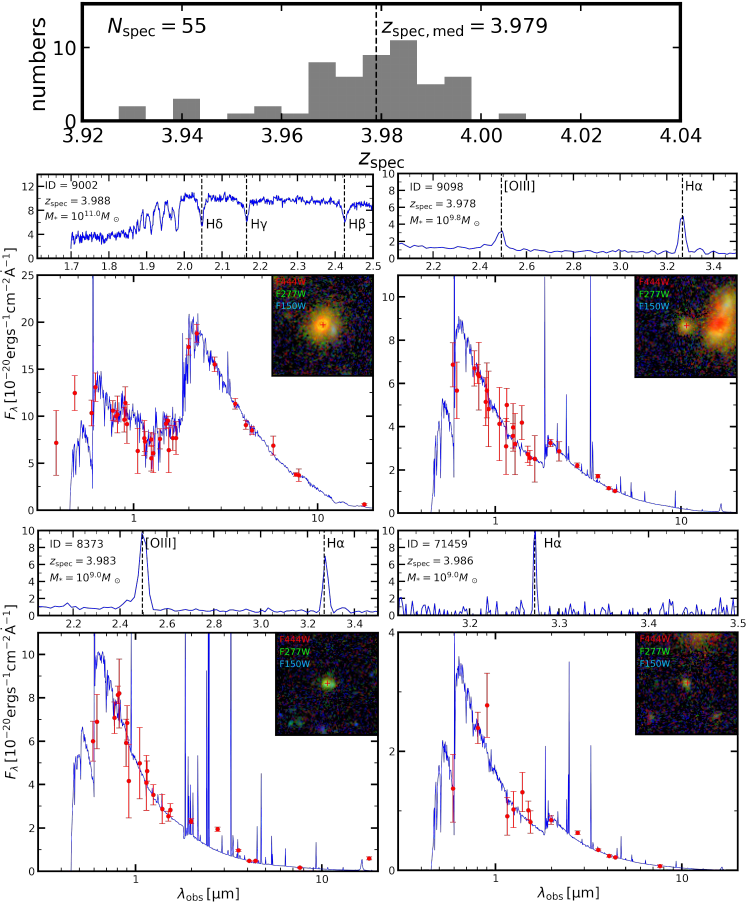}
\caption{\label{Fig:zspec}\footnotesize \textbf{Example of Spectroscopically confirmed members of the Bigfoot.} The upper panel presents the redshift distribution of 55 spectroscopically confirmed galaxies. As examples, the lower panels show the JWST spectroscopy of four confirmed member galaxies in 0217A, their best-fit SED from Bagpipes, and the $4"\times4"$ RGB cutouts from the JWST/NIRCam images.}
\end{figure*}

Using our multi-wavelength catalog for PRIMER-UDS, we first identify an extremely dense area with 6 galaxies in an area with radius $R=6$". This overdensity (the core of subgroup 0217A) is spectroscopically confirmed at ${z_{{\rm{spec}}}} = 3.98$, and is remarkably close ($\sim 16.7$cMpc) to an overdensity of quiescent galaxies at ${z_{{\rm{spec}}}} = 3.99$ reported by \citet{Tanaka2024} (Hereafter the T24 overdensity). Based on these two overdensities, we then investigate the distribution of overdensities around $z\sim4$ in the entire PRIMER-UDS field with JWST observations. The left panel of Figure~\ref{Fig:density} shows the distribution of mass-weighted number overdensity ${\Sigma _{{\rm{\tilde n}}}} = {\rho _{{\rm{\tilde n}}}}/{\rho _{{\rm{\tilde n,field}}}} - 1$ of galaxies with $3.73 < {z_{{\rm{phot}}}} < 4.23$, where the mass-weighted number density is defined as
\begin{equation}
\label{eq_mass_weight}
{\rho _{{\rm{\tilde n}}}} \equiv \sum\limits_{i = 1}^n {\log ({M_{*,i}}} /{10^{7.5}}{M_ \odot })/\rm{area}. \footnote{The completeness limit of the JWST observations from PRIMER is $\sim {10^{8.5}}{M_ \odot }$ at $z=4$, and we use the threshold ${10^{7.5 = 8.5-1}}$ to provide reasonable mass weighting. }
\end{equation}
For each colored area in Figure~\ref{Fig:density} that is at least two times more dense than field level, we search the spectroscopic data from three sources: Firstly, we collect JWST spectroscopy from the DAWN JWST Archive (DJA, version 3.0), which combines all available JWST spectra in PRIMER-UDS including the RUBIES survey \citep{WangBJ2024}, JWST projects GO 4233 (PI: Anna De Graaff), and GO 2565 (PI: Karl Glazebrook). These JWST spectra are processed by \textit{MSAEXP} \citep{msaexp,degraaff2024,Heintz2024}. Secondly, the VANDELS survey \citep{Talia2023} provides VLT/VIMOS spectra for galaxies across $1<z<6.5$ in PRIMER-UDS. Thirdly, we use the Keck/MOSFIRE spectra \citep{Tanaka2024} for five quiescent galaxies in the T24 overdensity.

Based on this combined spectroscopic data set, we identify the proto-cluster PCL0217 (the Bigfoot) with 11 subgroups spectroscopically confirmed at ${z_{{\rm{spec}}}} = 3.93 - 4.00$ (including 9 at ${z_{{\rm{spec}}}} = 3.97 - 4.00$)(We notice that although 0217B and 0217C are close in Figure~\ref{Fig:density}, they are actually $\sim 15$cMpc away from each other based on their spectroscopic redshifts). This includes 3 subgroups at $z = 3.93-4.00$ resolved from the T24 overdensity, and all 11 subgroups are located within a $15.5{\rm{cMpc}} \times 37.0{\rm{cMpc}}$ area. Combining these 11 subgroups, the Bigfoot has 55 spectroscopically confirmed members with ${z_{{\rm{spec}}}} = 3.927 - 4.009$ and ${z_{{\rm{spec,med}}}} = 3.979$. As an example, the JWST spectra of 4 spectroscopically confirmed members of the densest and most massive group 0217A are shown in Figure~\ref{Fig:zspec}, while the detailed position and spectroscopic redshift of each subgroup are listed in Table~\ref{Tab:phy_properties}.

\subsection{Selection of member galaxies}\label{sec:stellarM}

In addition to the spectroscopically confirmed members, we select a mass-complete sample of member galaxies using the photometric redshifts. The photometric redshifts in our catalog have $\sigma_{\mathrm{NMAD}} = 0.017$ \citep{WangT2025}, which corresponds to the $1\sigma$ uncertainty $\sigma (z=3.98) = 0.085$. Then, within a circular area with radius $R = 500$ kpc for each subgroup, we select all galaxies that have $\lvert z_{\rm phot} - 3.98 \rvert < 3\sigma$ ($3.73<z_{\mathrm{phot}}<4.23$) as members of the Bigfoot, which yields a sample of 755 member galaxies. Compared with the density of field galaxies at $3.73<z_{\mathrm{phot}}<4.23$ in PRIMER-COSMOS, the fraction of interlopers in the eight subgroups (0217A-H) covered by the PRIMER survey should be $\sim 26\%$($\sim 10\%$) in 500 (100)pkpc. The sample of member galaxies in these 8 subgroups is complete at $\log ({M_{\rm{*}}}/{M_ \odot }) > 8.5$ \citep{WangT2025}. On the other hand, for the three T24 overdensities (0217I-K) that only have the shallower JWST data from the BEACON survey \citep{Morishita2025}, the interloper fraction can be higher ($\sim 36\%$) and the sample of small galaxies ($\log ({M_{\rm{*}}}/{M_ \odot }) <9.0$) can be incomplete. This incompleteness is corrected using the best-fit SMF from Figure~\ref{Fig:SMF} before our analysis for the number densities and halo masses. 

We then perform SED fitting with Bagpipes \citep{Carnall2018} to estimate the stellar masses and rest-frame colors. During the SED fitting with Bagpipes, the redshifts of all selected members within the Bigfoot are fixed at the spectroscopic redshift of each subgroup. We utilize the 2016 update of the stellar population synthesis model \citep{Bruzual2003} with stellar ages allowed to range from 0.03 to 10 Gyr and metallicity $Z/Z_\odot\in[0.01,2.5]$. We adopt a delayed exponential star‐formation history with timescale $\rm \tau \in[0.01,10]~Gyr$, apply the dust models \citep{Calzetti2000} separately to young and old populations (split at 0.01 Gyr) with $A_V \in[0,5]$, and add nebular emission \citep{Byler2017} with ionization parameter $\log U\in[-5,-2]$. Examples of the best-fit SED with Bagpipes are shown in Figure~\ref{Fig:zspec}.

\subsection{Estimations of halo masses}\label{sec:haloM}

Based on the stellar masses estimated by Bagpipes, we estimate the halo mass of each subgroup of the Bigfoot in the following ways: Firstly, we summarize the stellar mass of all member galaxies with ${M_{\rm{*}}} > {10^{8.5}}{M_{_ \odot }}$ in each subgroup, which can be converted to a total stellar mass down to ${10^7}~{M_ \odot }$ following their best-fit stellar mass function. Then, using the redshift-dependent stellar-to-halo mass relation (SHMR) considering both central and satellite galaxies \citep{Shuntov2022}, we can obtain a halo mass ${M_{\rm{h,a}}}$ for all subgroups. Secondly, also based on the total stellar masses, we use the relation between the halo masses and total stellar masses of the galaxy clusters at $z \sim 1$ \citep{vanderBurg2014} to obtain ${M_{\rm{h,b}}}$. Thirdly, for the main group 0217A, which already shows a compact and concentrated core, we follow the methods used by the NICE team \citep{Sillassen2024} to estimate its halo mass ${M_{\rm{h,c}}}$ using its overdensity (105 times more dense than field level) within the virial radius. Lastly, the best estimated halo masses are given as the mean $\log {M_{\rm{h}}}$ from different methods. These best estimated halo masses are listed in Table~\ref{Tab:phy_properties}, whose errors are dominated by the uncertainty of the relation between total stellar masses and halo masses. 
For the central subgroup 0217A, it can be considered as one single halo since its concentrated density profile is described well by an NFW profile (see Figure~\ref{Fig:density_profile}), and we are using the area within its virial radius (131pkpc) to estimate the halo mass. For the other 10 subgroups except 0217A, we remind that their massive member galaxies could have extended distributions with the distance between them being larger than the maximum possible virial radius. These subgroups actually have several sub-halos, and we estimate the total halo mass of these sub-halos within a circular area with $R = 500\rm{pkpc}$.

\begin{table*}[!tbh]\footnotesize
\centering
\begin{minipage}[center]{\textwidth}
\centering
\caption{Properties of the 11 confirmed subgroups of PCL0217 (the Bigfoot).\label{Tab:phy_properties}}
\begin{tabular}{lcccccccccc}
\hline\hline
ID  & RA  & Dec & ${z_{{\rm{spec}}}}$ & ${N_{{\rm{spec}}}}$ & ${N_{{\rm{total}}}}$ & ${M_{ * ,{\rm{total}}}}$ & ${M_{\rm{h,a}}}$ & ${M_{\rm{h,b}}}$ & ${M_{\rm{h,c}}}$ & ${M_{\rm{h,best}}}$\\
& deg & deg & & & & $\log {M_ \odot }$ & $\log {M_ \odot }$ & $\log {M_ \odot }$ & $\log {M_ \odot }$ & $\log {M_ \odot }$\\
\hline
0217A & 34.3412 & -5.2404 & 3.98 & 5 & 81 & $11.6_{ - 0.1}^{ + 0.1}$ & $12.8_{ - 0.3}^{ + 0.3}$ & $13.0_{ - 0.3}^{ + 0.3}$ & $13.1_{ - 0.3}^{ + 0.3}$ & $13.0_{ - 0.3}^{ + 0.3}$\\
0217B & 34.3685 & -5.3098 & 3.99 & 9 & 66 & $11.4_{ - 0.1}^{ + 0.1}$ & $12.7_{ - 0.5}^{ + 0.3}$ & $12.7_{ - 0.5}^{ + 0.3}$ & - & $12.7_{ - 0.5}^{ + 0.3}$\\
0217C & 34.3654 & -5.2944 & 3.97 & 1 & 44 & $11.4_{ - 0.1}^{ + 0.1}$ & $12.7_{ - 0.5}^{ + 0.3}$ & $12.8_{ - 0.5}^{ + 0.3}$ & - & $12.8_{ - 0.5}^{ + 0.3}$\\
0217D & 34.3897 & -5.0870 & 3.97 & 6 & 84 & $11.4_{ - 0.1}^{ + 0.1}$ & $12.7_{ - 0.5}^{ + 0.3}$ & $12.7_{ - 0.5}^{ + 0.3}$ & - & $12.7_{ - 0.5}^{ + 0.3}$\\
0217E & 34.4586 & -5.1320 & 3.97 & 7 & 92 & $11.0_{ - 0.1}^{ + 0.1}$ & $12.5_{ - 0.5}^{ + 0.3}$ & $12.1_{ - 0.5}^{ + 0.3}$ & - & $12.3_{ - 0.5}^{ + 0.3}$\\
0217F & 34.3616 & -5.1494 & 3.98 & 9 & 86 & $10.9_{ - 0.1}^{ + 0.1}$ & $12.4_{ - 0.5}^{ + 0.3}$ & $12.0_{ - 0.5}^{ + 0.3}$ & - & $12.2_{ - 0.5}^{ + 0.3}$\\
0217G & 34.3327 & -5.1653 & 3.94 & 7 & 100 & $11.2_{ - 0.1}^{ + 0.1}$ & $12.5_{ - 0.5}^{ + 0.3}$ & $12.4_{ - 0.5}^{ + 0.3}$ & - & $12.4_{ - 0.5}^{ + 0.3}$\\
0217H & 34.4184 & -5.2964 & 3.98 & 7 & 108 & $11.1_{ - 0.1}^{ + 0.1}$ & $12.5_{ - 0.5}^{ + 0.3}$ & $12.3_{ - 0.5}^{ + 0.3}$ & - & $12.4_{ - 0.5}^{ + 0.3}$\\
0217I & 34.4437 & -5.3412 & 4.00& 2 & 32 & $11.2_{ - 0.1}^{ + 0.1}$ & $12.6_{ - 0.5}^{ + 0.3}$ & $12.4_{ - 0.5}^{ + 0.3}$ & - & $12.5_{ - 0.5}^{ + 0.3}$\\
0217J & 34.3954 & -5.3600 & 3.93 & 1 & 36 & $10.9_{ - 0.1}^{ + 0.1}$ & $12.4_{ - 0.5}^{ + 0.3}$ & $11.9_{ - 0.5}^{ + 0.3}$ & - & $12.2_{ - 0.5}^{ + 0.3}$\\
0217K & 34.3845 & -5.3875 & 3.99 & 1 & 26 & $10.9_{ - 0.1}^{ + 0.1}$ & $12.4_{ - 0.5}^{ + 0.3}$ & $11.9_{ - 0.5}^{ + 0.3}$ & - & $12.1_{ - 0.5}^{ + 0.3}$\\
\hline
\end{tabular}
\end{minipage}
\end{table*}

\section{Results}\label{sec:results}

Using the abundant spectroscopic data and a mass-complete selection of high redshift galaxies enabled by the deep JWST data in PRIMER-UDS \citep{Dunlop2021,Morishita2025}, we found a massive protocluster PCL0217 with 11 confirmed subgroups at $z \sim 3.98$ and present the distribution of these overdensities in Figure~\ref{Fig:density}. All of these 11 subgroups are located in an $14.6{\rm{cMpc}} \times 37.0{\rm{cMpc}}$ projected area in the sky and 49.0${\rm{cMpc}}$ (20.9${\rm{cMpc}}$ for the 9 subgroups except 0217G and 0217J) along the line of sight, which can be identified as one single protocluster by the friends-of-friends (FoF) method \citep{WangK2021}. Combining these subgroups, PCL0217 consists of 755 member galaxies, including 55 galaxies spectroscopically confirmed at ${z_{{\rm{spec}}}} = 3.927 - 4.009$ with $z_{\rm{spec,med}} = 3.979$. According to the spatial distribution of these 11 subgroups, we refer to it as the Bigfoot protocluster. Using the deep JWST/NIRCam and MIRI observations from the JWST/PRIMER \citep{Dunlop2021} and JWST/BEACON \citep{Morishita2025} surveys, we obtained accurate measurements of the total stellar masses and halo masses of the subgroups in the Bigfoot, which are listed in Table~\ref{Tab:phy_properties}. All of the 11 subgroups have ${M_{\rm{h}}} > {10^{12.1}}{M_{_ \odot }}$, and the most massive protocluster core already reaches a halo mass ${M_{\rm{h}}} = {10^{13.0}}{M_{_ \odot }}$.

\subsection{The Bigfoot as a progenitor of local galaxy clusters}

\begin{figure*}[!htbp]
\centering
\includegraphics[width=0.40\textwidth]{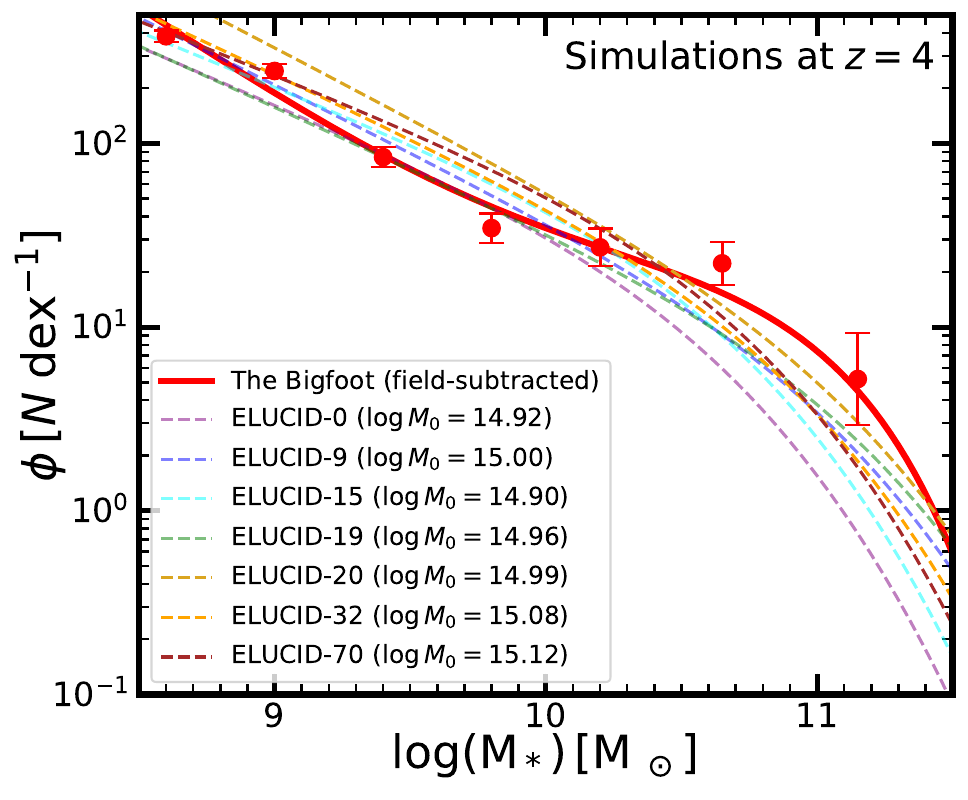}
\includegraphics[width=0.40\textwidth]{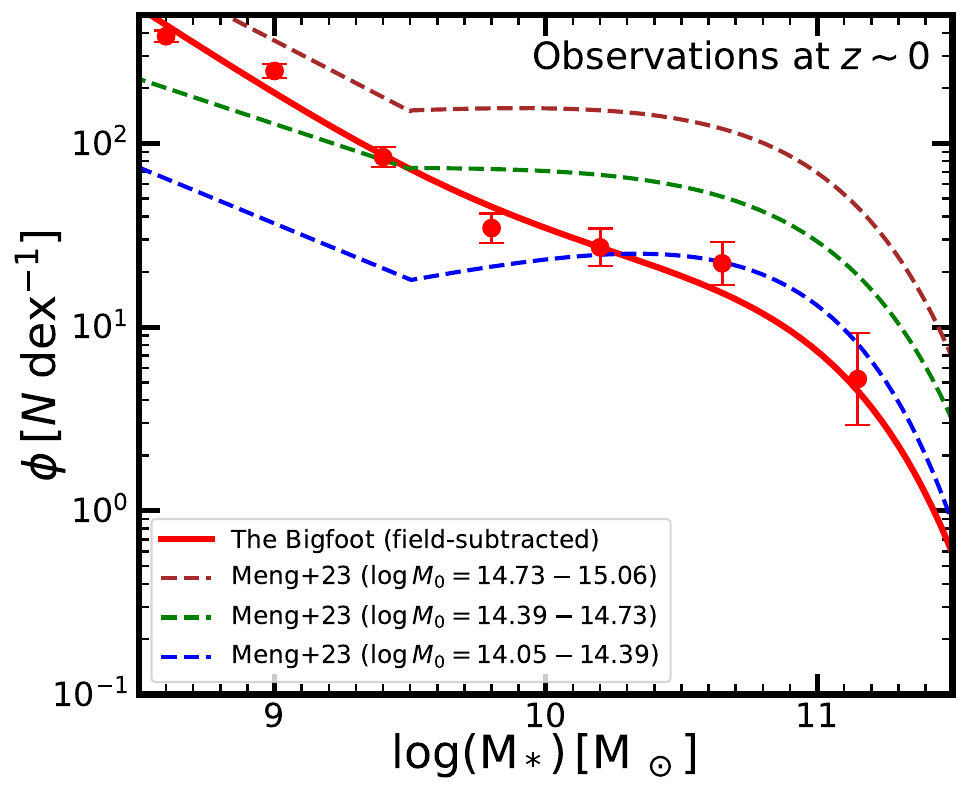}
\caption{\label{Fig:elucid}\footnotesize \textbf{The total SMF of the Bigfoot compared with massive clusters with ${M_0} \sim {10^{15}}{M_ \odot }$. } The left figure shows that the Bigfoot has a comparable number of galaxies and occupied volume to simulated progenitors of Coma-type massive galaxy clusters, where the red line shows the total SMF of the 9 subgroups of PCL0217 at ${z_{{\rm{spec}}}} = 3.97 - 4.00$ subtracted by a predicted contribution from the field SMF. The dashed lines show the total SMF of the progenitors ($z = 4.04$) of the 7 most massive galaxy clusters (${M_0} \sim {10^{15}}{M_ \odot }$) from the ELUCID simulation \citep{WangHY2014}. The right panel compares the total SMF of the Bigfoot with the observed mass functions of red galaxies in local clusters from DESI and SDSS \citep{Meng2023}.}
\end{figure*}

To understand the relation between the Bigfoot and the galaxy clusters in the local universe, we compare our observational results with the progenitors of local clusters from the cosmological dark matter simulation ELUCID \citep{WangHY2014,WangHY2016}. As a constrained simulation, ELUCID reconstructed and evolved the initial density field to reproduce the observed cosmic web, the halo mass function, and the density field in the local universe. ELUCID makes it possible to follow any local cluster back to its high-redshift antecedents, quantify its mass accretion and environmental evolution, and thereby evaluate whether an observed $z \sim 4$ protocluster has the requisite large-scale overdensities to mature into a present-day galaxy cluster. In the ELUCID simulation, we select the seven most massive clusters at $z=0$ with viral masses ${M_0} = {10^{14.90 - 15.12}}{M_ \odot }$. The progenitors of these clusters are widely extended as large-scale overdensities with a volume $4975-24562$ cMpc$^3$ at the $z = 4.04$ snapshot. The stellar-to-halo mass relation (SHMR) is adopted from the UniverseMachine \citep{Behroozi2019} to associate stellar masses to the simulated galaxy-scale subhalos (including the small subhalos of satellite galaxies). The scatter of this SHMR is considered by convolving the best-fit SMF shown in  Appendix~\ref{appendix:SMF}.

To make a fair comparison between the Bigfoot and the ELUCID clusters, we measure the total SMF of the 9 subgroups at $3.97 < {z_{{\rm{spec}}}} < 4.00$ (0217G at $z=3.94$ and 0217J at $z=3.93$ are excluded), and the field SMF in the same projected area at $3.73<z<4.23$ is subtracted to avoid the contamination from field galaxies. Considering the projected distribution of these 9 subgroups in the Bigfoot ($14.6{\rm{cMpc}} \times 37.0{\rm{cMpc}}$) and a redshift range $3.97-4.00$ (20.9${\rm{cMpc}}$), it yields a total SMF of the Bigfoot within a volume $\sim 11300$ cMpc$^3$, which is within the range of the cluster progenitors from ELUCID. Figure~\ref{Fig:elucid} shows that the total SMF of the Bigfoot is consistent with the total SMF of the simulated cluster progenitors at $\log ({M_{\rm{*}}}/{M_ \odot }) < 10.2$, which indicates that the Bigfoot has a comparable number of satellite galaxies located in a comparable volume with these proto-clusters from ELUCID. Moreover, as shown in Figure~\ref{Fig:Mz}, the mass of the most massive halo in the Bigfoot is also consistent with the simulated progenitors of the massive (${M_{\rm{0}}} \gtrsim {10^{15}}{M_ \odot }$) galaxy clusters. These facts suggest that the Bigfoot can be considered as a progenitor of a massive "Coma"-type galaxy cluster with ${M_0} \gtrsim {10^{15}}{M_ \odot }$ at $z = 0$. Meanwhile, the number of massive galaxies in the Bigfoot is much higher than the simulated protoclusters from ELUCID, suggesting that the formation of the massive galaxies in high-density environments could be earlier and more efficient than the prediction of simulations based on subhalo abundance matching.

The total SMF of the Bigfoot is compared with the observed mass function of red galaxies in local clusters \citep{Meng2023} in the right panel of Figure~\ref{Fig:elucid}. Based on archaeology studies \citep{Thomas2010} and numerical simulations \citep[e.g.][]{Chiang2013,Henden2020,WangK2025}, protoclusters at $z\sim 4$ will keep forming their massive galaxies until $z\sim 1.5-2$, so the total SMF of the Bigfoot is much lower than local clusters with ${M_{\rm{0}}} \sim {10^{15}}{M_ \odot }$. However, even if we do not consider any further star formation and mergers, the number of massive galaxies in the Bigfoot is already comparable to local clusters with ${M_{\rm{0}}} \sim {10^{14.2}}{M_ \odot }$, which provides a solid lower limit of its descendant at $z = 0$. 

\subsection{Top-heavy SMF with excess of massive galaxies in massive halos}
\begin{figure*}[!htbp]
\centering
\includegraphics[width=1.0\textwidth]{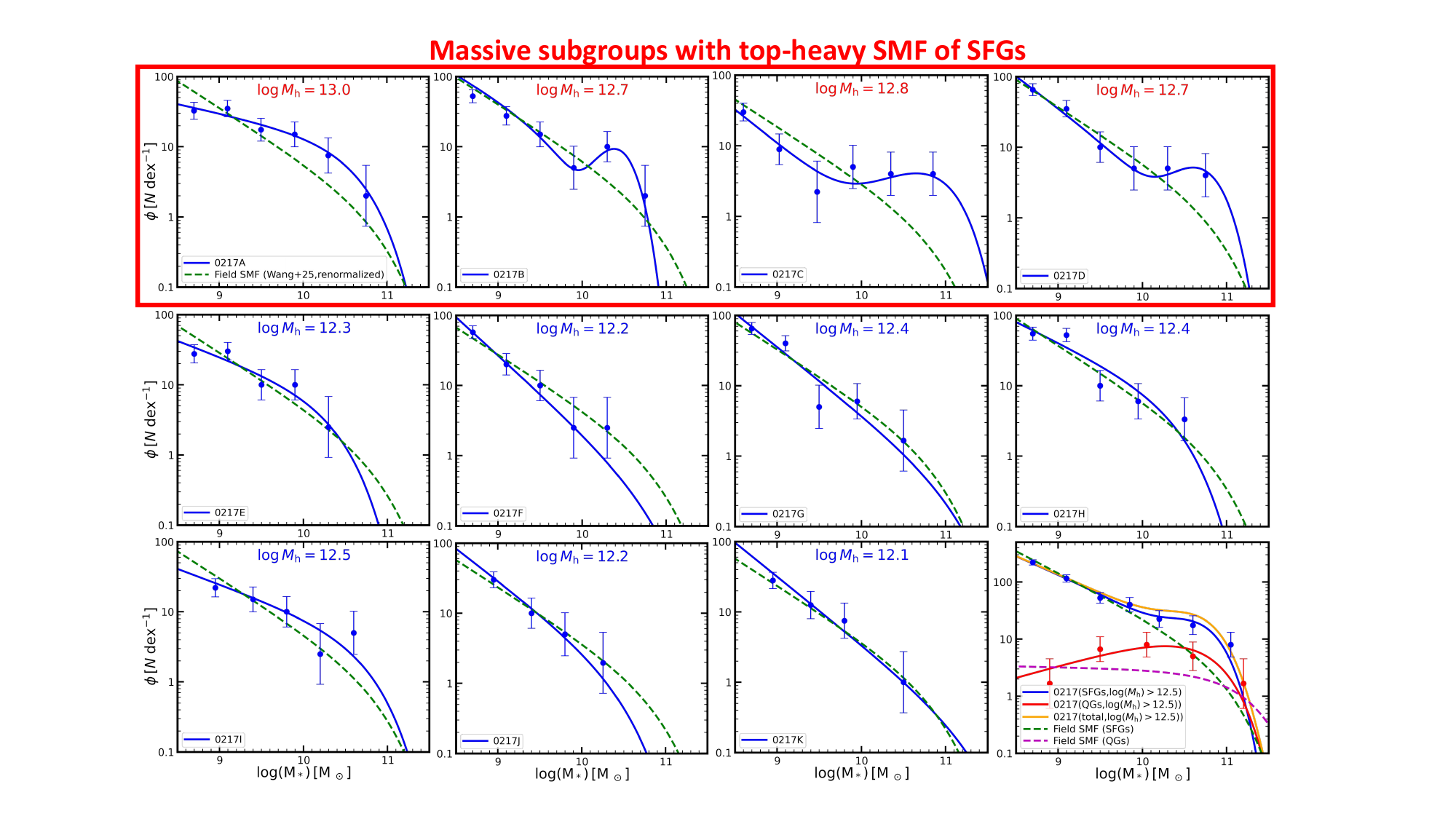}
\caption{\label{Fig:SMF}\footnotesize \textbf{Stellar mass functions with massive galaxies excess in the Bigfoot.} In the first 11 panels, the blue lines show the SMFs of SFGs in each subgroup of the Bigfoot, while the green dotted lines show the SMF of field galaxies at $3.5<z<4.5$ \citep{WangT2025}, which is based on the same catalog as this work and has been re-normalized to match the SMFs of the Bigfoot. All of the 4 massive subgroups show a top-heavy SMF of SFGs compared with the re-normalized field SMF and the less-massive subgroups, suggesting the accelerated formation of massive galaxies in massive overdensities. The last panel shows the total SMF of massive subgroups with $\log ({M_{\rm{h}}}/{M_ \odot }) > 12.5$, the blue line and the red line show the distribution of SFGs and QGs, respectively. The orange line shows the total SMF, which is the sum of the blue line and red line. The QGs, including both passive galaxies and post-starburst galaxies, are classified using an updated UVJ criterion \citep{Cutler2024}. Due to the limited number of QGs in each subgroup, we can only plot the total SMF of QGs instead of the SMF of QGs in each subgroup.}
\end{figure*}

Next, we measure the stellar mass functions for SFGs and quiescent galaxies (QGs) separately. The SFGs and QGs within the Bigfoot are classified using the rest-frame UVJ criterion \citep{Cutler2024} (We note that this updated UVJ criterion selects more small galaxies as quiescent post-starburst galaxies than the traditional UVJ selection \citep{Williams2009}). We identify 16 quiescent galaxies in the Bigfoot, 12/16 of which are located in the massive subgroups with ${M_{\rm{h}}} > {10^{12.5}}{M_{_ \odot }}$, including a massive (${M_{\rm{*}}} = {10^{11.0}}{M_{_ \odot }}$) quiescent galaxy in the center of the massive proto-cluster core 0217A (ID=9002 from Figure~\ref{Fig:zspec}, an ultra-massive QG previously reported by the JWST EXCELS survey\citep{Carnall2024}). This sample of quiescent galaxies yields a quiescent fraction of $17.9\%$ at ${M_{\rm{*}}} > {10^{10}}{M_{_ \odot }}$. For comparison, the quiescent fraction of the field galaxies in PRIMER-COSMOS based on the same UVJ criterion is only $6.45\%$, suggesting an enhanced quenching of the massive member galaxies in the Bigfoot.

Figure~\ref{Fig:SMF} presents the SMF of SFGs down to ${M_{\rm{*}}} > {10^{8.5}}{M_{_ \odot }}$ in each subgroup of the Bigfoot. Compared with the renormalized field SMF based on the same catalog for PRIMER-COSMOS, the SMFs of all four most massive subgroups in the Bigfoot (0217A-D) show an excess of massive SFGs at ${M_{\rm{*}}} > {10^{10}}{M_{_ \odot }}$, which is similar to the top-heavy SMF of cluster J1001 at $z = 2.51$ \citep{Sun2024}. On the contrary, the SMFs of the other smaller subgroups are similar to the field SMF. These excessive numbers of massive SFGs in massive halos show direct evidence of the enhanced star formation in protoclusters at high redshifts. On the other hand, in the local galaxy clusters, this top-heavy feature is only seen in the SMF of QGs or red galaxies \citep{Meng2023}, suggesting that many of these massive SFGs in protoclusters are likely to be quenched and evolve into the massive and red QGs observed in local clusters.

\subsection{The highly concentrated density profile of 0217A}
\label{sec:0217A}
\begin{figure*}[!htbp]
\centering
\includegraphics[width=0.45\textwidth]{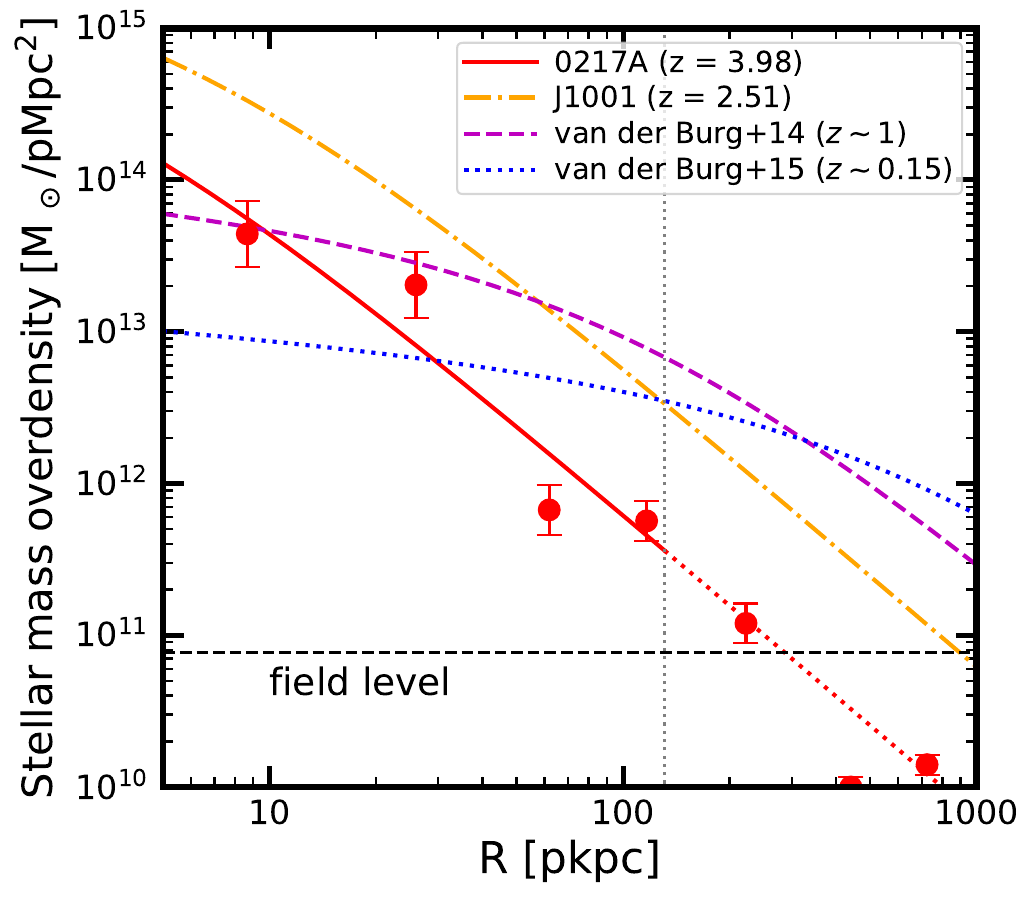}
\includegraphics[width=0.447\textwidth]{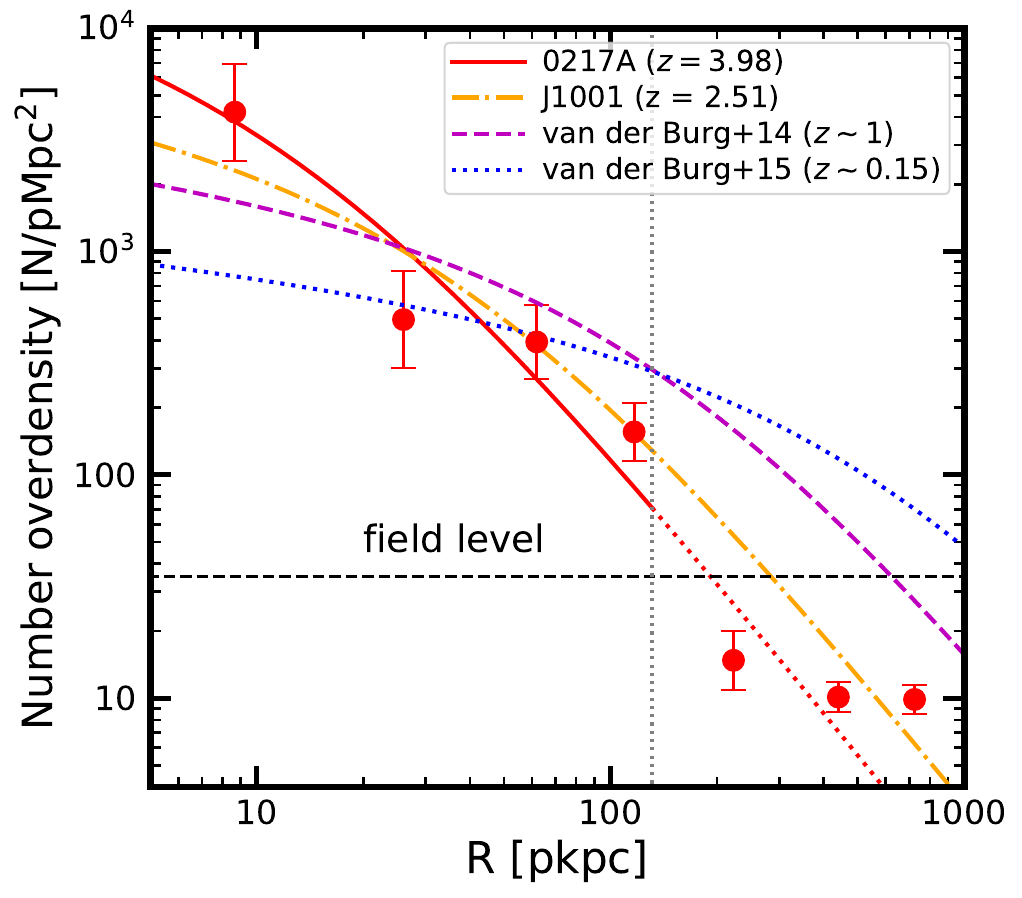}
\caption{\label{Fig:density_profile}\footnotesize \textbf{The stellar mass and number density profile of 0217A}: the left panel shows the projected stellar mass density profile, while the right panel shows the projected number density profile. The surface overdensities of 0217A are shown as the red lines, which have been subtracted by the field level (estimated as the density of field galaxies at $3.73<z<4.23$ in PRIMER-COSMOS). For comparison, the orange dash-dotted lines show the density profile of cluster J1001 at $z = 2.51$ \citep{Sun2024}, the purple dashed lines show the result of the GCLASS clusters at $z \sim 1$ \citep{vanderBurg2014}, and the blue dotted lines show the results of local clusters at $z \sim 0.15$ \citep{vanderBurg2015}. All of these archival results have been corrected to the density down to ${M_{\rm{*}}} = {10^{8.5}}{M_{_ \odot }}$ using their total SMF.}
\end{figure*}

Within the Bigfoot, the most massive group 0217A is a protocluster core that is already highly concentrated at a dense center with both a massive dusty star-forming galaxy detected by the ALMA SCUBA-2 UDS survey \citep{Stach2019}($870{\rm{\mu m}}$ continuum detection only, but it is merging with another confirmed member at ${z_{{\rm{spec}}}} = 3.978$) and a massive quiescent galaxy. The RGB color map of this region is shown in Figure~\ref{Fig:RGB}. For 0217A, we measure the projected profile of its stellar mass density and number density in Figure~\ref{Fig:density_profile}, during which we use all galaxies at ${M_{\rm{*}}} > {10^{8.5}}{M_{_ \odot }}$, and the most massive galaxy is excluded as in previous works \citep{vanderBurg2014,vanderBurg2015,Sun2024}.  After subtracting the density profile by the density of field galaxies, we find that the projected density profile of 0217A can be fitted by a projected Navarro-Frenk-White (NFW) profile \citep{Navarro1995,Bartelmann1996} with concentration ${c_{\rm{M}}} = 48_{ - 31}^{ + \infty }$ and ${c_{\rm{N}}} = 8.7_{ - 4.3}^{ + \infty }$.

To make a fair comparison between the density profile of 0217A and galaxy clusters at lower redshifts, we correct the archival density profiles down to ${M_{\rm{*}}} = {10^{8.5}}{M_{_ \odot }}$ using their total SMF. According to this comparison, the density of 0217A in its center ($R < 30$pMpc) is already comparable with these clusters at lower redshifts, but the density of 0217A in the outskirts is much lower, supporting an inside-out formation scenario of (proto)clusters \citep{vanderBurg2015,Sun2024} since $z = 4$. 

\section{Discussion}
\label{sec:discussion}

\subsection{Rareness of the Bigfoot and its cosmological implications}
\begin{figure*}[!htbp]
\centering
\includegraphics[width=0.40\textwidth]{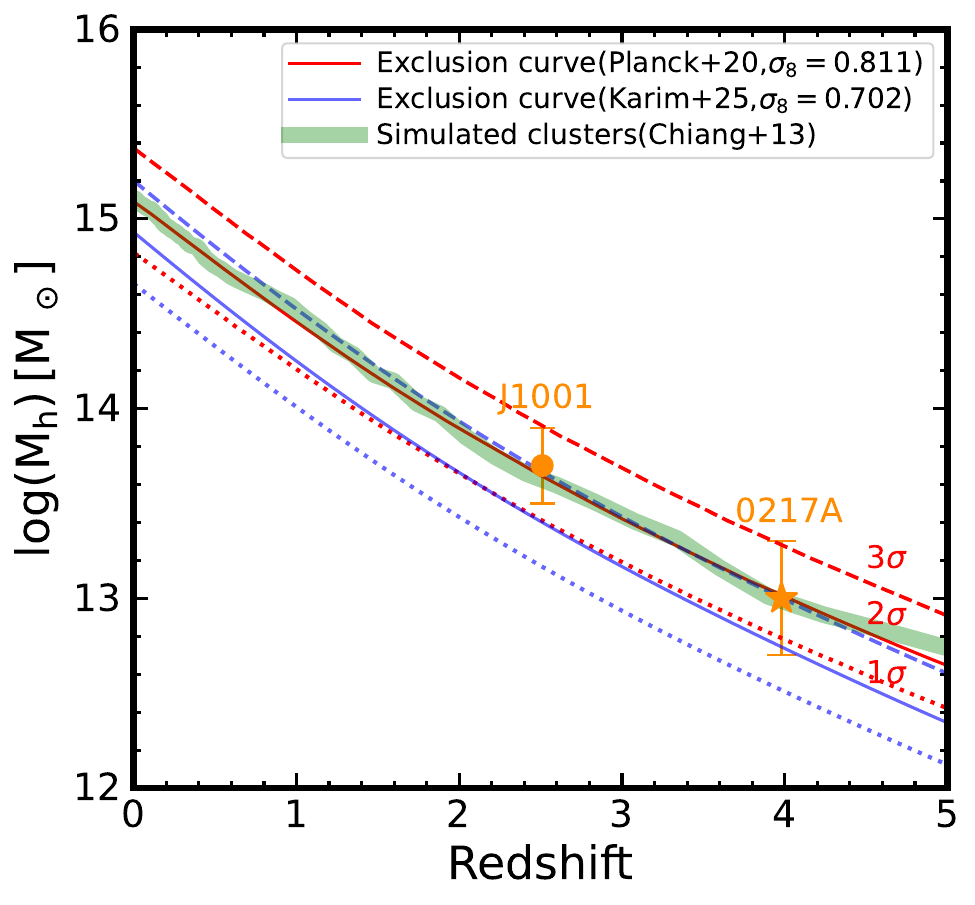}
\includegraphics[width=0.53\textwidth]{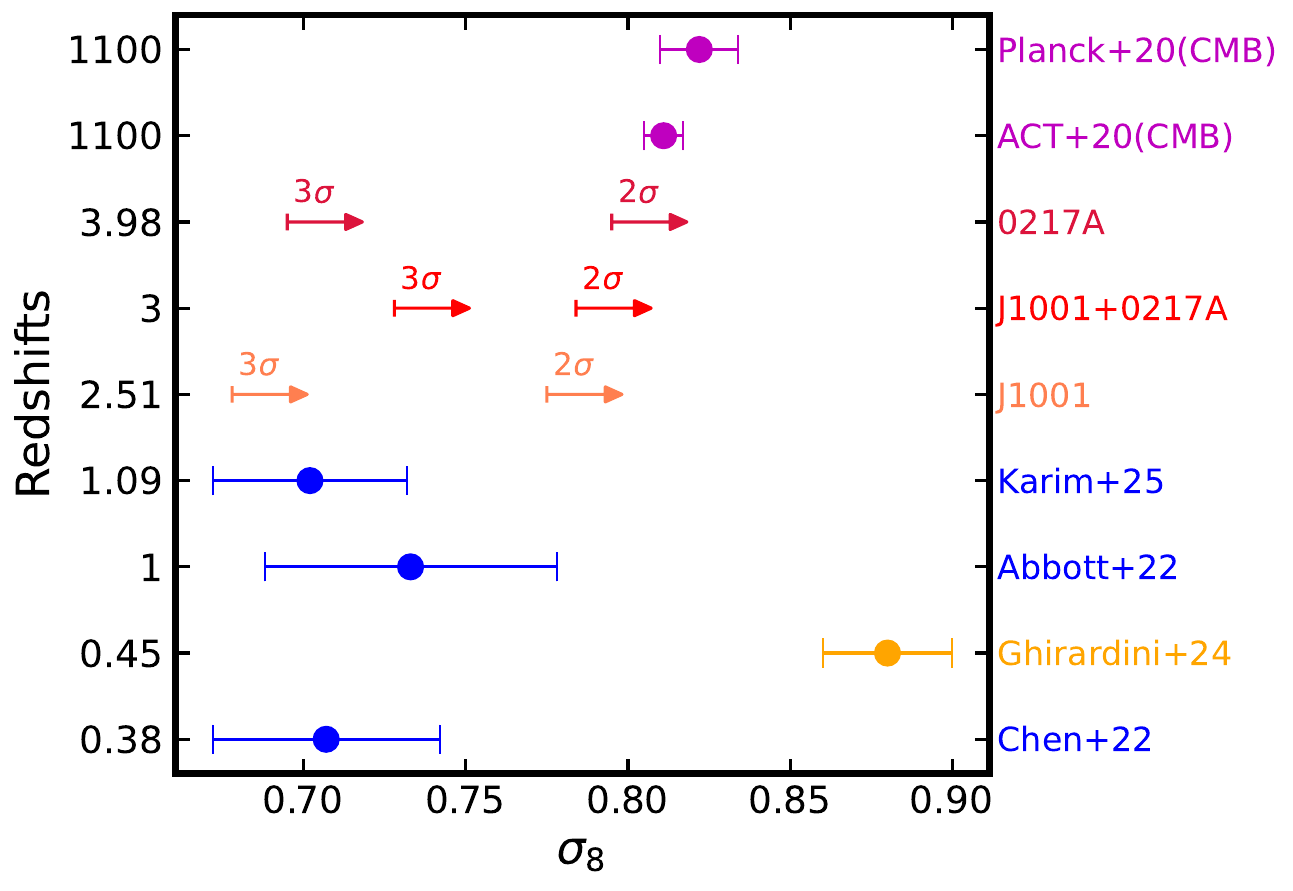}
\caption{\label{Fig:Mz}\footnotesize \textbf{The halo mass of 0217A compared with the constraint from cosmology and the simulated clusters.} In the left panel, the orange star shows the halo mass and redshift of 0217A. The red dotted, solid, and dashed lines show the 0.683 (1$\sigma$), 0.954 (2$\sigma$), and 0.997 (3$\sigma$) exclusion curves \citep{Harrison2013} in the 0.17deg$^2$ area under the Planck cosmology with ${\sigma _8} = 0.811 \pm 0.006$ and ${\Omega _{\rm{m}}} = 0.315 \pm 0.007$ \citep{Planck2020}. The blue lines show the exclusion curves under the cosmology measured by DESI \citep{Karim2025} with ${\sigma _8} = 0.702 \pm 0.030$ and ${\Omega _{\rm{m}}} = 0.285 \pm 0.010$. The green line shows the mass evolution of the most massive progenitor halo in the simulated massive clusters with ${M_0} > {10^{15}}{M_ \odot }$ \citep{Chiang2013}. In the right panel, the blue points show the ${\sigma _8}$ measured by the low redshift probes from BOSS \citep{Chen2022}, DES year 3 \citep{Abbott2022}, and the DESI Legacy Imaging Surveys \citep{Karim2025}, the orange point shows the ${\sigma _8}$ constrained by the local galaxy clusters from the SRG/eROSITA all-sky survey \citep{Ghirardini2024}, the arrows show the limits of ${\sigma _8}$ under which the Bigfoot, J1001, and the combination of them meet the 0.954 ($2 \sigma$) and 0.997 ($3 \sigma$) exclusions, and the magenta points show the ${\sigma _8}$ measured with CMB \citep{Aiola2020,Planck2020}.}
\end{figure*}

As a progenitor of ``Coma''-type massive galaxy cluster at $z=3.98$ located in the deep PRIMER-UDS field with limited available area, the Bigfoot provides an extremely rare opportunity to study the formation of massive galaxy clusters in the early universe. Notably, the densest protocluster core 0217A already reaches ${M_{\rm{h}}} = {10^{13.0}}{M_{_ \odot }}$ at $z = 4$. Figure~\ref{Fig:Mz} evaluates the rarity of this halo across the two PRIMER fields ($\sim 0.17$ deg$^2$) within $3.5<z<4.5$ by comparing it with the exclusion curves \citep{Harrison2013}. The rarity of a given cluster largely depends on ${\sigma _8}$, whose value is currently subject to a tension between high- and low-redshift cosmological probes. On one hand, using the cosmic microwave background (CMB) anisotropies at $z \sim 1100$, the Planck team \citep{Planck2020} reported ${\sigma _8} = 0.811 \pm 0.006$. On the other hand, the ${\sigma _8}$ measured from low-redshift probes can be much lower. For example, based on the galaxy clustering and weak lensing at low redshifts, the Dark Energy Survey \citep{Abbott2022} reported ${\sigma _8} = 0.733 \pm 0.045$. Moreover, by cross-correlating 27 million ELGs from the DESI Legacy Imaging Surveys \citep{Day2019}, ${\sigma _8}$ is measured to be ${\sigma _8} = 0.702 \pm 0.030$ \citep{Karim2025}. Figure~\ref{Fig:Mz} presents the results under both cosmologies from Planck \citep{Planck2020} and \citet{Karim2025}, showing that 0217A corresponds to a $\sim 2 \sigma$ fluctuation (0.929 exclusion) under the Planck cosmology but a $\sim 3 \sigma$ fluctuation (0.996 exclusion) when the lower ${\sigma _8}$ is adopted. 

Similar to the Bigfoot in PRIMER-UDS, a massive protocluster J1001 was also reported at $z = 2.51$ in the center of the COSMOS field \citep{WangT2016}. Using the velocity dispersion of member galaxies in J1001 and its total X-ray luminosity, the halo mass of J1001 can be estimated to be ${M_{\rm{h}}} = {10^{13.7}}{M_{_ \odot }}$, which is also consistent to a progenitor of massive galaxy cluster (${M_0} > {10^{15}}{M_ \odot }$) at $z = 2.51$. This halo mass of cluster J1001 within the 0.54deg$^2$ area covered by the COSMOS-Web survey \citep{Casey2023} is consistent with the 0.908 \citep{Planck2020} and 0.993 \citep{Karim2025} exclusion curves. Considering both 0217A and J1001 in a combined area with redshift range $2<z<4.5$, they agree with the 0.885 and 0.9994 exclusions under the two cosmologies. The fact that we can find a Coma progenitor in both of these two deep fields with limited areas shows a non-negligible tension with the low ${\sigma _8}$ reported by \citet{Karim2025}, and we prefer the Planck cosmology with ${\sigma _8} = 0.811 \pm 0.006$. On one hand, this tension of ${\sigma _8}$ could be explained by a ${\sigma _8}$ (or ${S_8}$) increasing with redshifts under the ${\rm{\Lambda CDM}}$ cosmology \citep{Adil2024,Akarsu2025}. On the other hand, these lower ${\sigma _8}$ values could be caused by biased observations, with several works also reporting high ${\sigma _8}$ using low-redshift probes. For example, the cosmology constrained by the local galaxy clusters at $0.1<z<0.8$ observed by the SRG/eROSITA all-sky survey \citep{Ghirardini2024} is ${\sigma _8} = 0.88 \pm 0.02$. This suggests that the cosmology probed by galaxy clusters across different redshifts is consistent with the measurement based on the CMB.

\subsection{Implications on the formation of the first massive clusters}
Based on a complete sample of member galaxies in the Bigfoot with deep JWST observations, we measure the SMFs for all subgroups and the density profile of the central halo in the Bigfoot. Compared with the SMF of field galaxies, the SMF of SFGs in all four massive subgroups in PLC0217 shows a top-heavy feature with too many massive galaxies that are star-forming in the massive halos. Meanwhile, the combined quiescent fraction in these massive subgroups is also elevated. These enhanced fractions of massive galaxies within massive halos suggest that the dense environment in the early universe significantly boosts the formation of massive cluster galaxies. One possible explanation for it is the feedback-free starbursts at cosmic dawn \citep{LiZ2024}, which leads to an enhanced baryon conversion efficiency at the high-mass end, and its prediction is consistent with the recent JWST observations \citep{WangT2025}. Meanwhile, the excess of massive substructures within a massive dark-matter halo in dense environments \citep{De-Lucia2004, Markos2024} could also help to produce the top-heavy SMF. 

The most massive subgroup 0217A in the Bigfoot is a protocluster core that has a concentrated density profile, which can be well described by an NFW profile \citep{Navarro1995}. This suggests that this core region might already be in the process of virialization. According to the comparison shown in Figure~\ref{Fig:density_profile}, the center region of 0217A is already as dense as the (proto)clusters at lower redshifts \citep{vanderBurg2014,vanderBurg2015,Sun2024}. By contrast, both the number and stellar mass density of 0217A in the outskirts are much lower. The combination of this highly concentrated density profile and the top-heavy SMFs of SFGs supports the inside-out and top-to-bottom formation scenario of the massive protoclusters at high redshifts, meaning that the massive central galaxies in protoclusters are formed ahead of the less massive galaxies in the outskirts \citep{Sun2024}.

\section{Conclusion}
\label{sec:conclusion}
In this work, we report the discovery of a massive protocluster PCL0217 (the Bigfoot) at $z=3.98$, which has 11 spectroscopically confirmed subgroups located in the deep PRIMER-UDS field. Using the UV to MIR multi-wavelength catalog based on deep JWST imaging and spectroscopic data in PRIMER-UDS, we present a completeness census of the member galaxies in the Bigfoot and study its global properties, including the SMFs and density profiles. Our main findings are listed as follows:

1. The large-scale expansion, total SMF, and the central halo mass of the Bigfoot are all consistent with the Coma progenitors from numerical simulations \citep{Chiang2013,WangHY2014}, suggesting that the Bigfoot will evolve into a massive Coma-type galaxy cluster with ${M_0} \gtrsim {10^{15}}{M_ \odot }$ at $z=0$.

2. Within the Bigfoot, all massive subgroups show enlarged fractions of massive galaxies. The most massive subgroup, 0217A, shows a highly concentrated density profile. These support the inside-out and top-to-bottom formation scenario of massive protoclusters at high redshift.

3. Combining the Bigfoot with another massive protocluster J1001 at $z=2.51$ in COSMOS \citep{WangT2016}, we argue that the presence of these two massive protoclusters located in two randomly placed JWST deepfields with limited available area strongly supports the high ${\sigma _8}$ value reported by \citet{Planck2020} rather than the lower ${\sigma _8}$ inferred from low-redshift probes \citep{Karim2025}.
 
\section{Acknowledgments}
This work was supported by the National Natural Science Foundation of China (Project No.12173017 and Key Project No.12141301), the National Key R\&D Program of China (grant no. 2023YFA1605600), the Scientific Research Innovation Capability Support Project for Young Faculty (Project No. ZYGXQNJSKYCXNLZCXM-P3), and the China Manned Space Program with grant no. CMS-CSST-2025-A04. Some of the data products presented herein were retrieved from the Dawn JWST Archive (DJA). DJA is an initiative of the Cosmic Dawn Center (DAWN), which is funded by the Danish National Research Foundation under grant DNRF140. The raw JWST imaging data are obtained from the Mikulski Archive for Space Telescopes (MAST) and are available at \href{http://dx.doi.org/10.17909/bf7a-bz25}{10.17909/bf7a-bz25}.

\facilities{JWST, VLT, Keck, HST, Subaru, UKIRT, Spitzer, CFHT}

\appendix
\restartappendixnumbering
\section{The RGB map of 0217A}
\label{appendix:RGB_0217A}
\begin{figure*}[!htbp]
\centering
\includegraphics[width=1\textwidth]{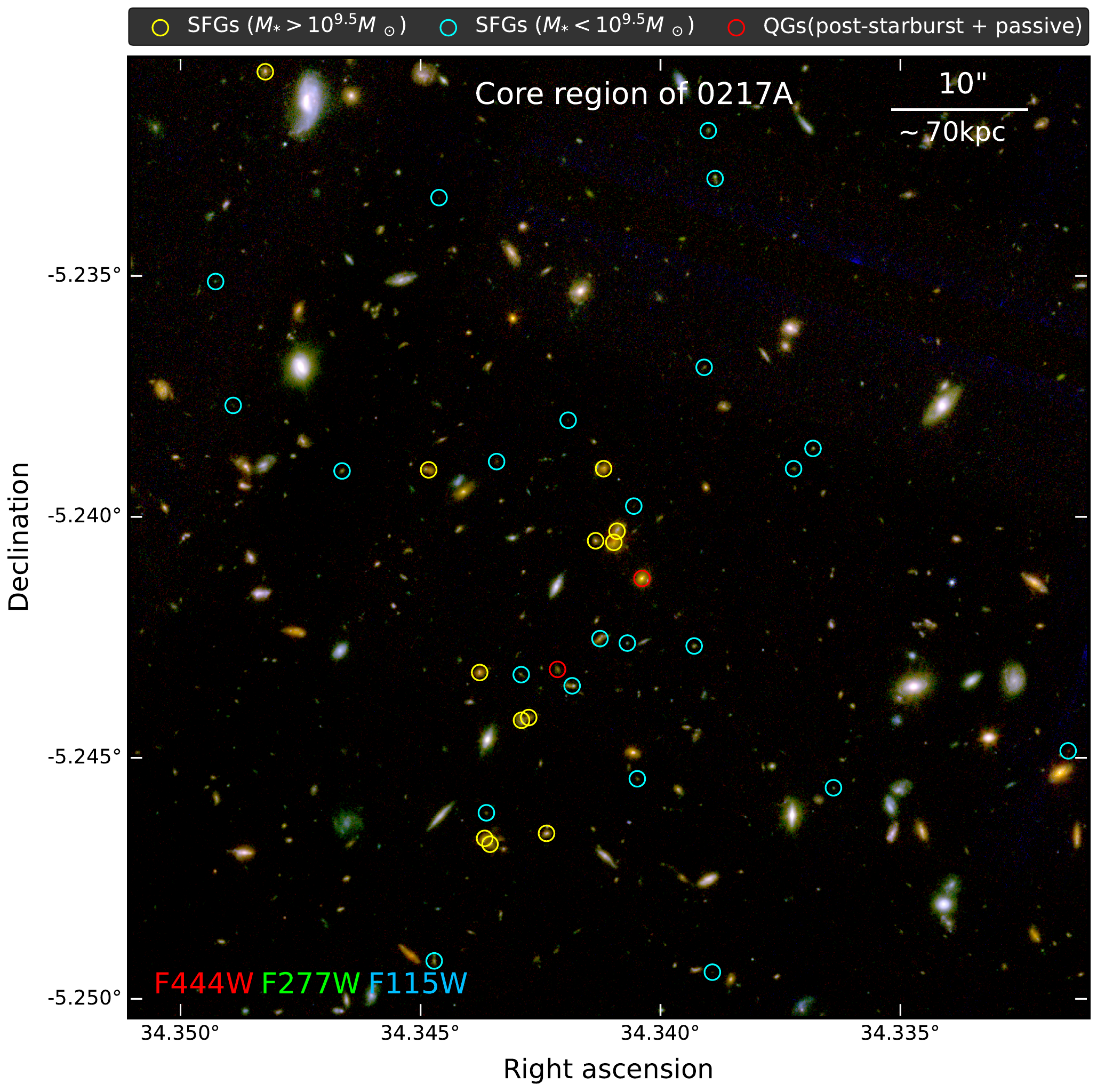}
\caption{\label{Fig:RGB}\footnotesize \textbf{The RGB composite color image of the most massive protocluster core (0217A).} The red, green, and blue channels correspond to the JWST/NIRCam F444W, F277W, and F150W filters, respectively. Member galaxies are overlaid as open circles: massive SFGs (${M_{\rm{*}}} > {10^{10}}{M_{_ \odot }}$) in yellow, low-mass SFGs (${M_{\rm{*}}} < {10^{9}}{M_{_ \odot }}$) in cyan, and quiescent galaxies (QGs, including both passive galaxies and post-starburst galaxies) in red. }
\end{figure*}

Figure~\ref{Fig:RGB} shows the RGB color map of the protocluster core 0217A with the selected member galaxies marked as open circles.

\section{The construction of SMF}
\label{appendix:SMF}
As described in \citet{Sun2024}, all SMFs are fitted by maximum-likelihood estimation (MLE), which is performed by minimizing the negative log likelihood with the Expectation Maximization algorithm \citep{Dempster1977}. With the MLE fitting, we can avoid the uncertainty caused by the arbitrary binning procedure. The range of the stellar mass used for MLE is $8.5 < \log ({M_{\rm{*}}}/{M_ \odot }) < 12.0$ for 0217A-H and $9.0 < \log ({M_{\rm{*}}}/{M_ \odot }) < 12.0$ for 0217I-K. The uncertainties of the best-fit parameters include both the Poisson error and the uncertainties of stellar masses.

For each SMF, we fit it using both the single Schechter function
\begin{equation}
\label{eq_3s}
\begin{aligned}
\Phi {\rm{d}}(\log M) = & \ln (10) \times \exp ( - {10^{\log M - \log {M^ * }}}) \\
& \times [\Phi _1^ * {({10^{\log M - \log {M^ * }}})^{{\alpha _1} + 1}}]{\rm{d}}(\log M),
\end{aligned}
\end{equation}
and the double Schechter function 
\begin{equation}
\label{eq_3}
\begin{aligned}
\Phi {\rm{d}}(\log M) = & \ln (10) \times \exp ( - {10^{\log M - \log {M^ * }}}) \\
& \times [\Phi _1^ * {({10^{\log M - \log {M^ * }}})^{{\alpha _1} + 1}}\\
 & + \Phi _2^ * {({10^{\log M - \log {M^ * }}})^{{\alpha _2} + 1}}]{\rm{d}}(\log M),
\end{aligned}
\end{equation}
Then, we determine which model is better using the Bayesian
Information Criterion (BIC). If $\Delta ({\rm{BIC}}) \equiv {\rm{BI}}{{\rm{C}}_{{\rm{single}}}} - {\rm{BI}}{{\rm{C}}_{{\rm{double}}}} > 0$, it suggests that this SMF can be better fitted by a double Schechter function. Otherwise, we provide the best-fit single Schechter function. Table~\ref{Tab:SMF_parameters} shows the best-fit parameters of the SMFs in this work. We caution that for most of the SMFs in this work, the quality of the single Schechter fitting and double Schechter fitting is comparable with $\Delta ({\rm{BIC)}} < 8 \sim 10$ \citep{Kass1995}.

\begin{table*}\footnotesize
\centering
\begin{minipage}[center]{\textwidth}
\centering
\caption{Best-fit parameters of the stellar mass functions. \label{Tab:SMF_parameters}}
\begin{tabular}{lcccccc}
\hline\hline
  & $\log ({M^ * })$  & $\Phi _{_1}$ & ${\alpha _1}$ &  $\Phi _{_2}$ & ${\alpha _2}$ & $\Delta ({\rm{BIC)}}$ \\
  & $({M_ \odot })$& $({\rm{de}}{{\rm{x}}^{ - 1}})$ & & $({\rm{de}}{{\rm{x}}^{ - 1}})$ & &  \\
\hline
Total$^a$ &  $10.95 \pm 0.07$ & $2.7 \pm 1.2$ & $-1.90 \pm 0.04$ &$9.9 \pm 2.5$ &$-1.06 \pm 0.27$ & 2.8\\
Total ($\log ({M_{\rm{h}}}/{M_ \odot }) > 12.5$, SFGs) &  $10.37 \pm 0.19$ & $8.1 \pm 2.6$ & $-1.63 \pm 0.06$ &$19.2 \pm 3.7$ &$0.5 \pm 1.2$ & 0.7\\
Total ($\log ({M_{\rm{h}}}/{M_ \odot }) > 12.5$, QGs) &  $10.65 \pm 0.09$ & $7.11 \pm 0.67$ & $-0.58 \pm 0.12$ & - & - & -5.8\\
0217A &  $10.59 \pm 0.10$ & $5.04 \pm 0.48$ & $-1.26 \pm 0.04$ & - & - & -2.7\\
0217B &  $9.78 \pm 0.28$ & $6.3 \pm 2.1$ & $-1.69 \pm 0.22$ &$0.85 \pm 0.43$ &$3.02 \pm 0.84$ & 3.1\\
0217C &  $10.70 \pm 0.24$ & $0.11 \pm 0.06$ & $-1.95 \pm 0.21$ &$4.7 \pm 1.6$ &$-0.06 \pm 0.81$ & 5.2\\
0217D &  $10.32 \pm 0.36$ & $1.07 \pm 0.58$ & $-1.89 \pm 0.13$ &$3.9 \pm 1.0$ &$1.0 \pm 1.1$ & 4.8\\
0217E &  $10.34 \pm 0.32$ & $2.81 \pm 0.84$ & $-1.45 \pm 0.13$ & - & - & -3.2\\
0217F &  $10.85 \pm 0.28$ & $0.11 \pm 0.05$ & $-2.09 \pm 0.20$ & - & - & -2.1\\
0217G &  $11.14 \pm 0.27$ & $0.14 \pm 0.06$ & $-1.96 \pm 0.15$ & - & - & -3.4\\
0217H &  $10.41 \pm 0.23$ & $2.76 \pm 0.87$ & $-1.58 \pm 0.12$ & - & - & -0.5\\
0217I &  $10.73 \pm 0.11$ & $1.85 \pm 0.39$ & $-1.44 \pm 0.10$ & - & - & -1.7\\
0217J &  $10.48 \pm 0.31$ & $0.56 \pm 0.20$ & $-1.91 \pm 0.15$ & - & - & -1.9\\
0217K &  $11.36 \pm 0.29$ & $0.07 \pm 0.02$ & $-1.96 \pm 0.14$ & - & - & -1.4\\
\hline
\end{tabular}
\begin{flushleft}
{\sc Note.} --- 
($a$) To recover the total SMF in Figure~\ref{Fig:elucid}, a field SMF with $\log ({M^ * }) = 10.94$, $\Phi _{} = 2.5$, and ${\alpha}=-1.77$ should be subtracted.
\end{flushleft}
\end{minipage}
\end{table*}

\bibliographystyle{aasjournal}
\bibliography{PCL0217_reference}
\end{document}